\documentclass[conference]{IEEEtran}
\usepackage{graphicx}
\usepackage{tabularx}

\usepackage[T1]{fontenc}
\usepackage[utf8]{inputenc}
\usepackage[english]{babel}

\usepackage{csquotes}
\usepackage{nicefrac}
\usepackage{afterpage}
\usepackage{needspace}
\usepackage[textsize=scriptsize]{todonotes}
\usepackage{booktabs}
\usepackage{tikz}
\usepackage{quantikz}
\usetikzlibrary{arrows,positioning} 
\usetikzlibrary{shapes.geometric, positioning}
\usetikzlibrary{fit, quotes}
\usetikzlibrary{shapes.geometric, positioning}
\usetikzlibrary{automata,arrows}
\usepackage{flushend}
\usepackage{hyperref}
\usepackage{soul}
\usepackage{amssymb}
\usepackage{csvsimple}
\usepackage{amsthm}
\usepackage{amsmath}
\usepackage{lipsum}
\usepackage{varwidth}
\usepackage{utfsym}
\usepackage{fontawesome5}
\usepackage{multicol}
\usepackage{listings}
\usepackage{adjustbox}
\usepackage{float}
\usepackage{placeins}
\usepackage{yquant}
\usepackage{pgfplots}
\pgfplotsset{compat=1.18}
\usepackage[detect-all=true]{siunitx}
\usepackage{etoolbox}
\usepackage{bbm}
\usepackage{balance}
\usepackage{changes}
\usepackage{makecell}
\usepackage{algorithm}
\usepackage{algpseudocodex}

\robustify\bfseries

\makeatletter
\let\MYcaption\@makecaption
\makeatother
\usepackage[font=footnotesize]{subcaption}
\makeatletter
\let\@makecaption\MYcaption
\makeatother

\newtheorem{example}{Example}

\usepackage[style=ieee, maxnames=9, minnames=6, date=year, doi=false, isbn=false, url=false, backend=biber, sortcites]{biblatex}
\AtEveryBibitem{
  \clearname{editor}
  \clearfield{series}
  \clearfield{isbn}
  \clearfield{issn}
  \clearfield{volume}
  \clearfield{number}
  \clearfield{pages}
  \clearfield{doi}
}

\newif\ifdoubleblind

\makeatletter
\newcommand{\redacted}[2][]{
  \ifdoubleblind
    \if\relax\detokenize{#1}\relax
      [redacted for double-blind review]%
    \else
      #1%
    \fi
  \else
    #2%
  \fi
}

\newcommand{\nothing}{}
\makeatother

\newcommand{\bx}{\mathbf{x}}
\newcommand{\bs}{\mathbf{y}}

\newcommand{\encodingNode}[3][]{\node[circle, draw,fill=cyan!30] (#2) at (#3) {};\node (#2_ONLY_TEXT) at (#3) {\tiny #1};}
\newcommand{\auxNode}[3][]{\node[circle, draw,fill=green!20] (#2) at (#3) {};\node (#2_ONLY_TEXT) at (#3) {\tiny #1};}

\usepackage{microtype}
\doubleblindfalse
\renewcommand{\baselinestretch}{0.86}

\hypersetup{
	pdftitle={Quantum Hardware-Efficient Selection of Auxiliary Variables for QUBO Formulations},                      %
	pdfsubject={Design, Automation \& Test in Europe (DATE)},                    %
	pdfauthor={Damian Rovara, Lukas Burgholzer, Robert Wille}    %
}

\begin{document}

\renewcommand*{\figureautorefname}{Fig.}
\renewcommand*{\sectionautorefname}{Section}
\renewcommand*{\subsectionautorefname}{Section}
\def\exampleautorefname{Example}

\title{Quantum Hardware-Efficient Selection of \\Auxiliary Variables for QUBO Formulations\redacted[\vspace*{-1.2cm}]{\vspace*{-0.2cm}}}

\author{
  \redacted[\nothing]{
\IEEEauthorblockN{Damian Rovara\IEEEauthorrefmark{1}\hspace*{1.5cm}Lukas Burgholzer\IEEEauthorrefmark{1}\IEEEauthorrefmark{2}\hspace*{1.5cm}Robert Wille\IEEEauthorrefmark{1}\IEEEauthorrefmark{3}\IEEEauthorrefmark{2}}
\IEEEauthorblockA{\IEEEauthorrefmark{1}Chair for Design Automation, Technical University of Munich, Germany}
\IEEEauthorblockA{\IEEEauthorrefmark{2}Munich Quantum Software Company GmbH, Garching near Munich, Germany}
\IEEEauthorblockA{\IEEEauthorrefmark{3}Software Competence Center Hagenberg GmbH (SCCH), Austria}
\IEEEauthorblockA{\href{mailto:damian.rovara@tum.de}{damian.rovara@tum.de}\hspace*{1.5cm}\href{mailto:lukas.burgholzer@tum.de}{lukas.burgholzer@tum.de}\hspace*{1.5cm}\href{mailto:robert.wille@tum.de}{robert.wille@tum.de}\\
\url{https://www.cda.cit.tum.de/research/quantum}}
\vspace*{-1.2cm}
  }
}

\maketitle

\begin{abstract}
The \emph{Quantum Approximate Optimization Algorithm} (QAOA) requires considered optimization problems to be translated into a compatible format.
A popular transformation step in this pipeline involves the quadratization of higher-order binary optimization problems, translating them into \emph{Quadratic Unconstrained Binary Optimization} (QUBO) formulations through the introduction of auxiliary variables.
Conventional algorithms for the selection of auxiliary variables often aim to minimize the total number of required variables without taking the constraints of the underlying quantum \mbox{computer---in particular,} the connectivity of its \mbox{qubits---into} consideration. 
This quickly results in interaction graphs that are incompatible with the target device, resulting in a substantial compilation overhead even with highly optimized compilers.
To address this issue, this work presents a novel approach for the selection of auxiliary variables tailored for architectures with limited connectivity.%
\phantom{x}By specifically constructing an interaction graph with a regular structure and a limited maximal degree of vertices, we find a way to construct QAOA circuits that can be mapped efficiently to a variety of architectures.
We show that, compared to circuits constructed from a QUBO formulation using conventional auxiliary selection methods, the proposed approach reduces the circuit depth by almost 40\%.
An implementation of all proposed methods is publicly available at \redacted{https://github.com/munich-quantum-toolkit/problemsolver}.
\end{abstract}

\section{Introduction}
\label{sec:introduction}

While large-scale quantum algorithms (such as Shor's~\cite{shor1994} or Grover's~\cite{grover1996} algorithm) still cannot be feasibly executed on state-of-the-art quantum computers, many experts in the field seek more tangible advantages through \mbox{utility-scale} algorithms compatible with \emph{Near-Term Intermediate Scale Quantum} (NISQ) devices.
Such algorithms are typically used to approximate solutions to optimization problems and focus on low-depth circuits that are employed in conjunction with classical \mbox{optimizers~\cite{cerezo2021, peruzzo2014,farhi2014, mcclean2016, kandala2017hardware}}.
A popular example for these \mbox{utility-scale} algorithms is the \emph{Quantum Approximate Optimization Algorithm}~(QAOA)~\cite{farhi2014}.
This algorithm takes advantage of a hybrid quantum-classical approach to find the ground state of a Hamiltonian operator representing the desired optimization problem, which, in turn, leads to its optimal solution.

To solve an optimization problem using QAOA, the problem must first be reformulated into a compatible format, such as the \emph{Quadratic Unconstrained Binary Optimization} (QUBO) formalism~\cite{kochenberger2004, kochenberger2014,lucas2014}.
This involves the construction of a binary cost function~$Q(\bx)$ with~$\bx \in \{0, 1\}^N$ consisting only of \emph{linear terms} of the form $c_i\bx_i$ or \emph{quadratic terms} of the form $c_{ij}\bx_i\bx_j$ for some real coefficients $c_i$ and $c_{ij}$.
In cases where the original cost function requires additional constraints or \mbox{higher-order} terms, additional variables must typically be introduced to remain within the rules of the QUBO formalism~\cite{glover2019, ayodele2022, verma2022}.
Once such a QUBO cost function has been constructed, it can be translated into a QAOA circuit in a straightforward manner that can then be executed on a quantum computer.

However, one key concern of NISQ devices lies in the compilation problem: 
As quantum computers typically only support a subset of quantum gates~\cite{miller2011, barenco1995, sriluckshmy2023} and do not allow arbitrary connectivity among all qubits~\cite{holmes2020,zulehner2019,wille2023,hopf2026}, a quantum circuit defined in a device-agnostic way must first be compiled to the target architecture before it can be executed.
This compilation procedure introduces overhead to the circuit that often makes it perform substantially worse than the ideal device-agnostic circuit would suggest.
Therefore, the task of efficiently compiling quantum circuits is an important area of research in the field of quantum computing design automation~\cite{burgholzer2026, wille2022}.

For the specific use case of compiling QAOA circuits for specific architectures, a myriad of previous work has already shown the potential improvements that can be achieved by considering the compilation problem for such a specialized scenario~\cite{alam2020,zhu2024,schmidbauer2025,ji2025}.
Similarly, previous work on the task of constructing QUBO formulations was able to automate the generation of QUBO cost functions from a variety of input specifications in an efficient manner~\cite{volpe2024, rovara2024,qubovert,docplex,nash2020}.
This work, however, attempts to combine these two approaches by taking device considerations already into account at the time of constructing the QUBO formulation.
This is achieved by selecting auxiliary variables in such a way that the structure of interactions between the variables remains regular in a way that can be mapped efficiently to the target machine.
While the number of qubits required to implement QAOA circuits from these hardware-efficient QUBO cost functions rises, the depth of these circuits is lower than conventionally created~QAOA~circuits, as overheads introduced by compilation are reduced. 
Our evaluations show a rising improvement as the considered problem size increases, with an average depth reduction of~39.2\% at problems with 16 variables.

The remainder of this work is structured as follows:
\autoref{sec:background} provides an overview on the construction of QUBO formulations and how QAOA circuits are typically generated from them.
\autoref{sec:motivation} then highlights the issues of this hardware-agnostic approach and instead motivates a solution for the hardware-efficient construction of QUBO formulations.
The algorithms proposed to construct such QUBO formulations and compile them to hardware-compatible QAOA circuits in the ideal case are then proposed in \autoref{sec:embedded-qubo-generation}, whereas \autoref{sec:special-cases} illustrates how special cases can be handled efficiently as well.
Finally, the results obtained during the evaluation of the proposed methods are summarized in \autoref{sec:evaluation} before \autoref{sec:conclusion} concludes this work.

\section{Background}
\label{sec:background}

This section briefly reviews the underlying concepts of QUBO problems and QAOA.

\subsection{QUBO Formulations}

To optimize a binary cost function using QAOA, the cost function is commonly first translated to the QUBO formalism\mbox{~\cite{kochenberger2004, kochenberger2014,lucas2014}}.
In particular, this requires the reduction of all higher-order terms to quadratic terms by introducing auxiliary variables, a procedure also known as quadratization~\cite{rosenberg1975, verma2020, anthony2017, schmidbauer2025, schmidbauer2024,schmidbauer2024b}.
Each auxiliary variable encodes a product of two variables \mbox{$\bs_i = \bx_{i_1}\bx_{i_2}$}.
By substituting all occurrences of this product by the corresponding auxiliary variable, the order of each updated term is reduced by~1.
The introduction of this auxiliary variable also requires an additional penalty term~\mbox{$P_i(\bx_a, \bx_b) = c_{P_i} \left ( \bx_a\bx_b - 2\bx_a\bs_{i} - 2\bx_b\bs_{i} + 3\bs_i \right )$} to be added to the cost function.
This penalty term ensures that the total cost cannot be reduced by violating the equality constraint \mbox{$\bs_i = \bx_a\bx_b$}.
To incentivize its minimization, each penalty term is multiplied by a constant cost factor $c_{P_i}$ that has to be chosen sufficiently large, depending on the full QUBO cost function.
Repeatedly applying these substitutions allows any higher-order unconstrained binary optimization problem to be reduced to a QUBO cost function.

\begin{example}
\label{ex:slack}
    Consider the following binary optimization function to be approximated using QAOA:
    $$C(\bx) = \bx_1 \bx_2 \bx_3 \bx_4$$
    This cost function, together with its general extension to $N$ variables $C_N(\bx) = \prod_i^N \bx_i$, will serve as running examples for this work.
    To translate this binary optimization problem into the QUBO formalism, we substitute (sub-)products of two variables with newly introduced auxiliary variables to reduce the order of the expression.
    An efficient way to achieve this is to select commonly occuring products first with the goal of minimizing the number of additionally introduced variables.
    For the considered expression, we may choose $\bs_1 = \bx_1 \bx_2$ and $\bs_2 = \bx_3 \bx_4$, resulting in the new cost function:
    $$Q(\bx, \bs) = \bs_1 \bs_2 + P_1(\bx_1, \bx_2) + P_2(\bx_3, \bx_4)$$
    with two additional variables $\bs_i$ and two \emph{penalty terms}~$P_i$ defined as above, that account for the substitution of the auxiliary variables.
    The corresponding penalty factors were both chosen as $c_{P_i} = 1$. 
\end{example}

The process of transforming general optimization problems into the QUBO format is a popular area of research.
Existing solutions provide approaches to find space-efficient QUBO formulations for higher-order optimization problems as well as for problems that include additional constraints~\cite{volpe2024, rovara2024,qubovert,docplex}.

\subsection{Optimizing QUBO Problems with QAOA}

Once a QUBO cost function is generated, algorithms such as QAOA~\cite{farhi2014, blekos2024} can be used to find its optimal variable assignment.
As a \emph{variational quantum algorithm} (VQA)~\cite{cerezo2021}, QAOA consists of two components: a \emph{parametrized quantum circuit} for the preparation of a state as well as a \emph{classical optimizer} that updates the circuit parameters to find the optimal variable assignment~\cite{bonet-monroig2023, powell1998}.

QAOA requires the QUBO cost function to be translated to a \emph{cost Hamiltonian} $H_C$.
The goal of the circuit is then to find the ground state of this Hamiltonian which corresponds to the assignment resulting in the minimal cost.
In practice, this is achieved by applying the operator $e^{-i\gamma H_C}$ to the uniform superposition $H^{\otimes n}\ket{0}$ for some parameter $\gamma$.
This is then followed by the application of $e^{-i\beta H_M}$ with some other parameter $\beta$, representing an evolution towards the \mbox{\emph{mixer Hamiltonian} $H_M$~\cite{fuchs2022, govia2021}}, which typically consists of a set of \texttt{X} rotations on each qubit.
These cost and mixer layers can be applied to the circuit repeatedly, utilizing new parameters $\gamma_i, \beta_i$ for each repetition layer.
While, often, a single application of the layers is already enough to yield results, it has been shown that each repetition further increases the accuracy of the algorithm~\cite{farhi2014}.
The general process of starting with a QUBO cost function and constructing the QAOA cost layer from it is illustrated in the following example.

\begin{figure*}
\centering
\includegraphics*[width=0.95\textwidth]{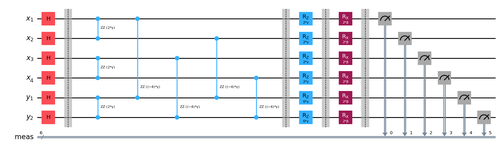}
\vspace{-1.5em}
\caption{The QAOA circuit for the optimization problem introduced in \autoref{ex:slack} using one repetition layer.}
\label{fig:circuit}
\vspace{-0.5cm}
\end{figure*}

\begin{example}
\label{ex:qaoa}
    Returning to the QUBO cost function introduced in \autoref{ex:slack}, the corresponding QAOA circuit requires 6 qubits:
    4 qubits represent the initial problem variables $\bx_i$, while the remaining two qubits represent the auxiliary variables~$\bs_j$.
    All qubits are initially set to a uniform superposition by applying an $H$ gate to each qubit.
    We then map the cost function to the cost layer in the QAOA circuit construction by adding an $\texttt{rzz}(2c_{ij}\gamma)$ gate for each quadratic term where~$c_{ij}$ is the term's coefficient in the cost function.
    Furthermore, each individual qubit also requires a $\texttt{rz}(2b_i\gamma)$ gate, \mbox{where $b_i = -2c_i - \sum_jc_{ij}$} is introduced by the translation from a binary cost function to a Hamiltonian.
    Here, $c_i$ represents the coefficient of each variable's linear term in the QUBO function.
    The resulting circuit is illustrated in \autoref{fig:circuit}.
\end{example}

The resulting QAOA circuit can then be used to find the expectation value of $H_C$ by repeatedy running the circuit and measuring the qubits.
A classical optimizer can use the measurement results to tune the parameters $\gamma_i,\beta_i$ until an assignment is found that yields an adequately low cost value.
This assignment can then be translated back directly to the assignment that minimizes the original cost function $C(\bx)$.

\section{Motivation and General Idea}
\label{sec:motivation}

This section describes the typical workflow of mapping QAOA circuits to quantum computers.
In the process, it highlights common challenges and limitations of the approach, and motivates how an architecture-aware selection of auxiliary variables during the construction of QUBO formulations can mitigate and resolve them.  

\subsection{Considered Problem}

Due to the limitations of state-of-the-art quantum computers, the depth of quantum circuits is a crucial criterion for their usefulness.
Not only do disproportionately deep circuits require more time for their execution, consuming valuable limited resources, but each additional gate has a negative impact on the circuit's probability of success due to gate errors and noise.
Unfortunately, the connectivity between hardware qubits is limited on many currently explored quantum computer architectures, such as superconducting systems.
Thus, additional gates (such as \texttt{SWAP}s) need to be inserted into the original circuit so that it conforms to the device's topology.
This typically leads to substantial increases in circuit depth.

\begin{figure}
\centering
\includegraphics[width=0.35\textwidth]{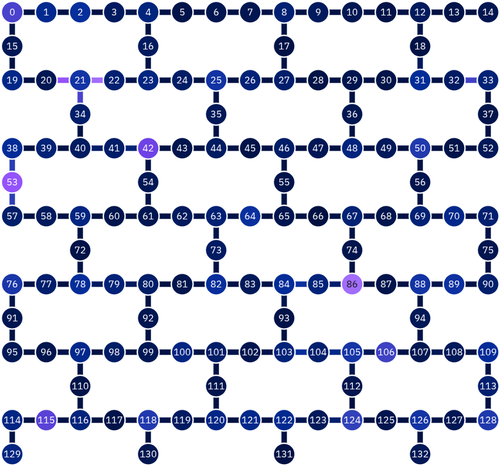}
\vspace{-0.5em}
\caption{The coupling map of the 133-qubit \texttt{ibm\_torino} superconducting quantum computer based on IBM's \emph{Heron} processor type.}
\label{fig:device}
\vspace{-1.5em}
\end{figure}

\begin{example}
\autoref{fig:device} shows the coupling map of the 133-qubit superconducting quantum computer \texttt{ibm\_torino} provided by the IBM Quantum Cloud platform~\cite{IBMQuantum}.
It employs a \emph{heavy-hex} topology, which is characterized by a grid of 12-qubit blocks that are arranged in a brick wall pattern.
In this topology, the maximum degree of nodes in the coupling map is 3, with a majority of qubits only being connected to 2 neighbors.
However, considering once again the binary cost function introduced in \autoref{ex:slack} and its corresponding QAOA circuit constructed in \autoref{ex:qaoa}, it is easy to see that the circuit is not directly compatible with this coupling map:
As the circuit in \autoref{fig:circuit} requires two directly connected qubits with a degree of 3, it must be modified to fit the device topology.
Specifically, this circuit requires at least 2 \texttt{SWAP} gates to be compatible with \texttt{ibm\_torino}, that each have to be further decomposed into the native gates supported by the device~\footnote{IBM's latest generation of quantum computers natively supports \texttt{id}, \texttt{x}, \texttt{sx}, \texttt{rz}, \texttt{rx}, \texttt{cz}, and \texttt{rzz}  gates. Based on that gateset, a SWAP can be decomposed to a set of 3 \texttt{cz} gates and 6 \texttt{sx} gates for a total depth of 6.}.
\end{example}

While finding the optimal circuit mapping to minimize the total depth is a computationally hard problem~\cite{boteaComplexityQuantumCircuit2018}, various optimization techniques have already been proposed for this task~\cite{wille2023,zulehner2019, siraichi2018, fosel2021, wagner2023, li2019, guo2025}.
Nonetheless, the impact of the increased depth caused by this procedure may greatly reduce the efficiency of the entire QAOA algorithm, as even the addition of a low number of \texttt{SWAP} gates may already increase the final circuit size considerably.
This effect only gets worse with larger circuits and more complex cost functions. 

While several approaches have been proposed to tackle this compilation overhead for QAOA circuits~\cite{alam2020,zhu2024,montanez-barrera2025,ji2025}, none of the existing approaches can really overcome the underlying problem that any created QUBO cost function might already be a bad fit for the architecture.
However, the process of constructing QUBO formulations from general binary optimization problems offers some degree of freedom in choosing auxiliary variables.
Crucially, this decision should already take into account the hardware's constraints in order to minimize the subsequent compilation overhead.

\subsection{General Idea}

Due to the commutative property of \texttt{rzz} and \texttt{rz} gates, the order of individual gates within a cost layer in the resulting circuit does not matter.
Therefore, we can focus all optimization efforts only on the interactions between variables and, in extension, the qubits that represent them.
To this end, we illustrate QUBO formulations as undirected graphs, where each vertex represents a variable in the expression and each edge between two vertices indicates that the product of the corresponding variables is a term in the full cost function.
Blue vertices represent the original problem variables and are labeled with $\bx_i$, while green vertices represent auxiliary variables and are labeled with $\bs_i$.

\begin{figure}
    \centering
    \begin{subfigure}{0.2\textwidth}
        \centering
        \begin{tikzpicture}
            \encodingNode[$\bx_1$]{x1}{0,0}
            \encodingNode[$\bx_2$]{x2}{1, -1}
            \encodingNode[$\bx_3$]{x3}{2,2}
            \encodingNode[$\bx_4$]{x4}{3,1}
            \auxNode[$\bs_1$]{y1}{1.0,0.0}
            \auxNode[$\bs_2$]{y2}{2,1}

            \draw (x1) -- (x2);
            \draw (x1) -- (y1);
            \draw (x2) -- (y1);
            \draw (x3) -- (x4);
            \draw (x3) -- (y2);
            \draw (x4) -- (y2);
            \draw (y1) -- (y2);
        \end{tikzpicture}
        \caption{$C(\bx) = \bx_1\bx_2\bx_3\bx_4$}
        \label{fig:interaction-graph-naive-4}
    \end{subfigure}
    \begin{subfigure}{0.2\textwidth}
        \centering
        \begin{tikzpicture}
            \encodingNode[$\bx_1$]{x1}{-1.0,2.25}
            \encodingNode[$\bx_2$]{x2}{-0.5,2.5}
            \auxNode[$\bs_1$]{y1}{-0.5,2.0}

            \encodingNode[$\bx_3$]{x3}{ 1.0,2.25}
            \encodingNode[$\bx_4$]{x4}{ 0.5,2.5}
            \auxNode[$\bs_2$]{y2}{ 0.5,2.0}

            \auxNode[$\bs_5$]{y5}{ 0.0,1.5}

            \encodingNode[$\bx_5$]{x5}{-1.0,-0.25}
            \encodingNode[$\bx_6$]{x6}{-0.5,-0.5}
            \auxNode[$\bs_3$]{y3}{-0.5,0.0}

            \encodingNode[$\bx_7$]{x7}{ 1.0,-0.25}
            \encodingNode[$\bx_8$]{x8}{ 0.5,-0.5}
            \auxNode[$\bs_4$]{y4}{ 0.5,0.0}

            \auxNode[$\bs_6$]{y6}{ 0.0,0.5}

            \draw (y1) -- (y2);
            \draw (y1) -- (y5);
            \draw (y2) -- (y5);
            \draw (y3) -- (y4);
            \draw (y3) -- (y6);
            \draw (y4) -- (y6);
            \draw (y5) -- (y6);

            \draw (x1) -- (x2);
            \draw (x1) -- (y1);
            \draw (x2) -- (y1);

            \draw (x3) -- (x4);
            \draw (x3) -- (y2);
            \draw (x4) -- (y2);

            \draw (x5) -- (x6);
            \draw (x5) -- (y3);
            \draw (x6) -- (y3);

            \draw (x7) -- (x8);
            \draw (x7) -- (y4);
            \draw (x8) -- (y4);
        \end{tikzpicture}
        \caption{$C_8(\bx) = \bx_1\bx_2...\bx_8$}
        \label{fig:interaction-graph-naive-8}
    \end{subfigure}
    \caption{The interaction graph constructed from the QUBO formulation proposed in \autoref{ex:slack}, as well as its extension to 8 variables, following the strategy of repeatedly replacing all sub-products of two variables by a new auxiliary variable until only a quadratic cost function is left.}
    \label{fig:interaction-graph-naive}
    \vspace{-1.5em}
\end{figure}
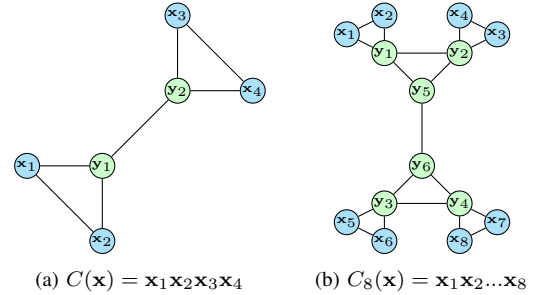

\begin{example}
\autoref{fig:interaction-graph-naive-4} illustrates the interaction graph constructed for the QUBO cost function introduced in \autoref{ex:slack}.
It is clearly visible how each edge in the graph directly translates to an \texttt{rzz} gate in the circuit in \autoref{fig:circuit}.
Comparing this interaction graph with the coupling map shown in \autoref{fig:device} demonstrates that it clearly cannot be mapped directly to the \texttt{ibm\_torino} device.
However, the interaction graph complexity rises even further when including additional variables in the initial cost function.
\autoref{fig:interaction-graph-naive-8} shows the interaction graph obtained when using the same approach of selecting auxiliary variables on a product of 8 binary variables.
Generally, as the number of variables grows, we similarly expect the interaction graph to become more complex.
Such graphs may contain vertices with drastically increasing degrees, leading to a similar rise in the required additional gates introduced by the compilation procedure.
\end{example}

By further investigating the nature of the interaction graph, we see that each auxiliary variable $\bs_k$ introduced to replace the product $\bx_{k_1}\bx_{k_2}$ leads to a triangle connecting the three variables in the interaction graph.
This shows how the selection of auxiliary variables can be leveraged to influence the required connections between qubits.
In particular, by selecting each new auxiliary variable $\bs_{i+1}$ in such a way that the previously added auxiliary variable $\bs_i$ is a factor of the replaced term, the resulting interaction graph will always be a chain of such triangles.
This way, even if the number of variables increases, the same chain-of-triangles structure will be retained consistently, allowing qubit mapping strategies to be proposed for that regular structure, rather than the less predictable graphs shown in \autoref{fig:interaction-graph-naive}. 

\begin{figure}
    \centering
    \begin{subfigure}{0.2\textwidth}
        \centering
        \begin{tikzpicture}[scale=0.8]
            \encodingNode[$\bx_4$]{x4}{0,0}
            \auxNode[$\bs_1$]{y1}{1,1}
            \auxNode[$\bs_2$]{y2}{2,2}
            \encodingNode[$\bx_1$]{x1}{2.5,2.5}

            \encodingNode[$\bx_3$]{x3}{0.1,0.9}
            \encodingNode[$\bx_3$]{x2}{1.1,1.9}

            \draw[red, ultra thick] (x4) -- (x3);
            \draw (x4) -- (y1);
            \draw[red, ultra thick] (x3) -- (y1);

            \draw[red, ultra thick] (y1) -- (x2);
            \draw (y1) -- (y2);
            \draw[red, ultra thick] (x2) -- (y2);

            \draw[red, ultra thick] (y2) -- (x1);
        \end{tikzpicture}
        \caption{$C(\bx) = \bx_1\bx_2\bx_3\bx_4$}
        \label{fig:interaction-graph-new-4}
    \end{subfigure}
    \begin{subfigure}{0.2\textwidth}
        \centering
        \begin{tikzpicture}[scale=0.5]
            \encodingNode[$\bx_8$]{x8}{0,0}
            \auxNode[$\bs_1$]{y1}{1,1}
            \auxNode[$\bs_2$]{y2}{2,2}
            \auxNode[$\bs_3$]{y3}{3,3}
            \auxNode[$\bs_4$]{y4}{4,4}
            \auxNode[$\bs_5$]{y5}{5,3}
            \auxNode[$\bs_6$]{y6}{6,2}
            \encodingNode[$\bx_1$]{x1}{7,1}

            \encodingNode[$\bx_7$]{x7}{0.1,0.9}
            \encodingNode[$\bx_6$]{x6}{1.1,1.9}
            \encodingNode[$\bx_5$]{x5}{2.1,2.9}
            \encodingNode[$\bx_4$]{x4}{3.1,3.9}
            \encodingNode[$\bx_3$]{x3}{4.9,3.9}
            \encodingNode[$\bx_2$]{x2}{5.9,2.9}

            \draw (x8) -- (y1);
            \draw (y1) -- (y2);
            \draw (y2) -- (y3);
            \draw (y3) -- (y4);
            \draw (y4) -- (y5);
            \draw (y5) -- (y6);
            \draw[red, ultra thick] (y6) -- (x1);

            \draw[red, ultra thick] (x8) -- (x7);
            \draw[red, ultra thick] (y1) -- (x7);
            \draw[red, ultra thick] (y1) -- (x6);
            \draw[red, ultra thick] (y2) -- (x6);
            \draw[red, ultra thick] (y2) -- (x5);
            \draw[red, ultra thick] (y3) -- (x5);
            \draw[red, ultra thick] (y3) -- (x4);
            \draw[red, ultra thick] (y4) -- (x4);
            \draw[red, ultra thick] (y4) -- (x3);
            \draw[red, ultra thick] (y5) -- (x3);
            \draw[red, ultra thick] (y5) -- (x2);
            \draw[red, ultra thick] (y6) -- (x2);

        \end{tikzpicture}
        \caption{$C_8(\bx) = \bx_1\bx_2...\bx_8$}
        \label{fig:interaction-graph-new-8}
    \end{subfigure}
    \caption{The interaction graph constructed by selecting auxiliary variables to form chains of triangles. 
    The red lines indicate the ordering in which the variables should be mapped to a path of qubits on the target device.}
    \label{fig:interaction-graph-new}
    \vspace{-1.5em}
\end{figure}
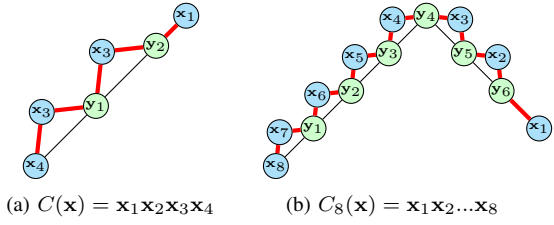

\begin{example}
Returning to the initial cost function defined in \autoref{ex:slack}, we propose an alternative selection of auxiliary variables.
We begin by substituting the product of any pair of variables, such as $\bs_1 = \bx_3 \bx_4$.
Then, we introduce the second auxiliary variable $\bs_2$ to substitute a product that includes $\bs_1$, such as $\bs_2 = \bx_2 \bs_1$.
This results in a new QUBO formulation that can be translated to the interaction graph illustrated in \autoref{fig:interaction-graph-new-4}.
Clearly, this graph includes two triangles ($\overline{\bs_1\bx_3\bx_4}$, and $\overline{\bs_2\bx_2\bs_1}$), corresponding to the two auxiliary variables.
Furthermore, by once again extending the approach to 8 variables in \autoref{fig:interaction-graph-new-8}, we see how the structure consistenly expands to a chain of 6 triangles for the 6 required auxiliary variables.
Any additional variables added to the product cost function $\prod_i^N \bx_i$ will result in an additional triangle in the interaction graph in the same consistent and regular manner.
\end{example}

While the resulting circuits still typically cannot be directly executed on devices with a limited connectivity, the regular structure of the circuits allows to derive application-specific placement and routing strategies.
By mapping the variables onto a line of qubits that support nearest neighbor connections according to the path highlighted in red in \autoref{fig:interaction-graph-new}, any interaction graph following this structure can be implemented by a circuit with constant depth.
In the following, we further investigate how such a circuit can be constructed in detail and what changes have to be applied to also support cost functions that cannot be represented by a single chain of triangles.

\section{Hardware-Efficient QUBO Generation}
\label{sec:embedded-qubo-generation}

Continuing from the general idea proposed above, this section illustrates in more detail how auxiliary variables can be selected in a hardware-efficient manner even in more complex cases.
It then shows how the resulting structure of auxiliary variables can be leveraged to efficiently compile QAOA circuits for specific architectures.

\subsection{Hardware-Efficient Auxiliary Variable Selection}
\label{sec:slack-selection}

To construct a QUBO formulation from higher-order binary cost functions, all terms of order 3 or higher must be reduced by adding auxiliary variables to the cost function.
Each auxiliary variable $\bs_i$ may substitute a product of two binary variables taken either from the original problem variables $\bx$ or the previously added auxiliary variables $\bs$.

To construct an interaction graph that consists of a maximal chain of triangles, each newly selected auxiliary variable $\bs_{i+1}$ must replace the previously introduced auxiliary variable $\bs_i$ as well as a problem variable $\bx_j$ that has not yet been subsituted by an auxiliary variable.
To this end, we propose a greedy algorithm that chooses the most commonly occurring $\bx_j$ to increase the likelihood of being able to further extend the chain.
Using the resulting substitution, it transforms the original cost function $C(\bx)$ into the QUBO cost function~$Q(\bx, \bs)$. 

We begin by selecting the sub-product that appears most frequently among all terms in $C(\bx)$ to be substituted by the first auxiliary variable.
All occurrences of the sub-product in~$C(\bx)$ are then replaced by the newly introduced auxiliary variable and the penalty term required by it is added to the cost function.
We then iterate through all terms in the remaining cost function with order greater than 2, searching for the problem variable $\bx_i$ that most frequently appears in a product together with the newly introduced auxiliary variable.
This product is substituted by a new auxiliary variable and the corresponding penalty term is added once again.
This process is then repeated, iteratively reducing the order of the cost function by substituting a previous auxiliary variable $\bs_i$ with a new variable $\bs_{i+1}$.
If the procedure can be repeated until no more higher-order terms remain in the cost function, the resulting interaction graph contains a single chain of triangles, as shown in \autoref{fig:interaction-graph-new}.

\begin{example}
Consider the following initial cost function:
$$C(\bx) = \bx_1\bx_2\bx_3\bx_4\bx_5 + \bx_1\bx_2\bx_3\bx_4 + \bx_2\bx_3\bx_4$$
The most frequently appearing sub-products are $\bx_3\bx_4$, $\bx_2\bx_4$, and $\bx_2\bx_3$.
We first substitute $\bx_3\bx_4$ by $\bs_1$.
This changes the cost function to
$$C(\bx, \bs) = \bx_1\bx_2\bx_5\bs_1 + \bx_1\bx_2\bs_1 + \bx_2\bs_1 + P_1(\bx_3, \bx_4)$$
Now, the most frequent problem variable appearing in a product with $\bs_1$ is $\bx_2$.
Therefore, $\bs_2 = \bx_2\bs_1$, changing the cost function to
$$C(\bx, \bs) = \bx_1\bx_5\bs_2 + \bx_1\bs_2 + \bs_2 + P_1(\bx_3, \bx_4) + P_2(\bs_1, \bx_2)$$
Finally, only the first term of the cost function has an order higher than two, thus, all contained variables appear with the same frequency.
For this example, we select $\bs_3 = \bs_2\bx_5$, resulting in the final QUBO cost function
\begin{multline*}
        Q(\bx, \bs) = \bx_1\bs_3 + \bx_1\bs_2 + \bs_2 \\ + P_1(\bx_3, \bx_4) +  P_2(\bs_1, \bx_2)
          + P_3(\bs_2, \bx_5).   
\end{multline*}
This QUBO formulation results in an interaction graph with a chain of triangles structure consisting of 3 triangles, each originating from one of the penalty functions $P_i$.
\end{example}

The proposed algorithm constructs a hardware-efficient QUBO formulation in a variety of cases.
While it does not succeed in finding a single chain of triangles in all cases, these special cases can also be handled in an efficient manner, as discussed later in \autoref{sec:special-cases}.

\subsection{Compiling QAOA Circuits from Interaction Graphs}
\label{sec:compilation}

Once a binary cost function is translated to an interaction graph based on the rules formulated in \autoref{sec:slack-selection}, it can be translated to a QAOA circuit using a set of simple rules.

We start by following a path of vertices on the interaction graph, starting with an edge between the first two problem variables that were substituted by $\bs_1$.
We then continue the path, iteratively moving to each subsequent auxiliary variable~$\bs_{i+1}$ connected by edges through the problem variable that has an edge to both $\bs_i$ and $\bs_{i+1}$.
Examples for such paths are highlighted in \autoref{fig:interaction-graph-new}.

We then find a path of qubits on the coupling map with the same size.
While finding such a path is a hard problem on general graphs~\cite{karger1997}, the regular structure of many superconducting device topologies often allows such a path to be found efficiently. 
For the heavy hex topology, for instance, as shown in \autoref{fig:device}, a path can be found by starting at the lowest-index qubit that is only connected to one neighbor ($14$ in the example) and continuing to traverse the coupling map, always selecting the qubit with the closest index.
For \texttt{ibm\_torino} with a total number of 133 qubits, this allows for a path with a maximum length of 112.

Each variable of the path on the interaction graph is then mapped to the qubit at its corresponding index in the path through the coupling map.
A $H$ gate is applied to each qubit to prepare a superposition for the QAOA circuit.
To apply the cost Hamiltonian, we then use the following~algorithm:

\begin{enumerate}
    \item \emph{Even direct interactions}: Apply an $\texttt{rzz}$ gate to each~$q_{2i}, q_{2i+1}$ for each \mbox{$i \in [0, \lfloor |q|/2 \rfloor]$}.
    As the selection of qubits supports nearest neighbor interactions and none of the required gates intersect each other, all gates can be performed in parallel in a single layer.
    \item \emph{Odd direct interactions}: Apply an $\texttt{rzz}$ gate to each~$q_{2i-1}, q_{2i}$ for each \mbox{$i \in [1, \lfloor |q|/2 \rfloor]$}.
    Once again, all of these gates can be applied in a single layer.
    \item \emph{Even indirect interactions}: Apply an $\texttt{rzz}$ gate to each~$q_{4i}, q_{4i+2}$ for each \mbox{$i \in [0, \lfloor |q|/4 \rfloor]$}.
    As there is a gap of size 1 between each pair of targets, each \texttt{rzz} gate requires a \texttt{SWAP} gate to move the qubits together and a second \texttt{SWAP} gate to move them back. 
    Therefore, this step only requires a constant depth.
    \item \emph{Odd indirect interactions}: Apply an $\texttt{rzz}$ gate to each~$q_{4i - 2}, q_{4i}$ for each \mbox{$i \in [1, \lfloor |q|/4 \rfloor]$}.
    Due to the gaps between the targeted qubits, the circuit once again requires \texttt{SWAP} gates to move them together.
    In this case, however, no second \texttt{SWAP} gate is required to move the qubits back, as the cost function is now complete and requires no further 2-qubit gates\footnote{If further repetition layers are required, odd-numbered iterations should instead apply step 4 first in reverse, returning the qubits to the original layout, and then continue with the remaining steps from 1 to 5.}.
    \item \emph{Linear terms}: Finally, for all linear terms in the cost function, apply an $\texttt{rz}$ gate to the corresponding qubit, keeping in mind possible changes to the qubit layout introduced during the previous step.
\end{enumerate}

For each applied rotation gate, the corresponding angle has to be computed as $2c\gamma$, where $c$ is the coefficient of the corresponding term in the cost function and $\gamma$ is the QAOA parameter of the current repetition layer.
In total, this process results in a QAOA cost Hamiltonian with a constant depth.
On \texttt{ibm\_torino}, the decomposition of \texttt{SWAP} gates mentioned in \autoref{sec:motivation} results in a total depth of 23 for one application of the cost Hamiltonian. 
As all individual steps of the above algorithm can be performed in parallel, the circuit depth will not grow, even if the interaction graph increases in size.

\vspace{-0.1em}

\section{Handling Special Cases}
\label{sec:special-cases}

The approach proposed above provides a general strategy for the efficient selection of auxiliary variables.
However, two additional scenarios must be handled depending on the initial cost function.
While a single, regular chain of triangles can be mapped efficiently to a variety of different topologies, these anomalies require additional adaptations that may increase the circuit depth.
In the following, we will discuss how these special cases can be handled and how hardware-efficient QAOA circuits can be constructed from them.

\subsection{Extraneous Interactions}

Pre-existing quadratic terms in the cost function are not considered during the choice of auxiliary variables.
While some of them may still be substituted by auxiliary variables that consider the same products, most of these pre-existing quadratic terms are not expected to be influenced by the auxiliary variable selection.
Therefore, the corresponding interactions will remain part of the resulting interaction graph.

Each of these \emph{extraneous interactions} requires a corresponding \texttt{rzz} gate to be added to the end of the cost layer of the circuit.
As these qubits will typically not be connected in the coupling map, additional \texttt{SWAP} gates must be added accordingly.
For this process, we employ conventional \texttt{SWAP} insertion methods~\cite{wille2023,zulehner2019}.
These additional \texttt{SWAP} gates will further increase the circuit depth, depending on the amount of extraneous interactions.

\subsection{Additional Chains}
\label{sec:additional-chains}

\begin{figure}
    \centering
    \begin{subfigure}{0.15\textwidth}
        \centering
        \begin{tikzpicture}[scale=0.7]
            \encodingNode{A}{-4.95,0.25}
            \encodingNode{B}{-4.9,1.0}
            \auxNode{C}{-4.25,1.05}
            \encodingNode{D}{-4.2,1.8}
            \auxNode{E}{-3.55,1.85}

            \auxNode{F}{-4.5,-0.3}
            \auxNode{G}{-4.0,0.3}
            \auxNode{H}{-3.7,-0.3}
            \auxNode{I}{-3.3,0.3}
            \auxNode{J}{-2.9,-0.3}
            \auxNode{K}{-2.5,0.3}
            \auxNode{L}{-2.1,-0.3}

            \draw (A) -- (B);
            \draw (A) -- (C);
            \draw (B) -- (C);
            \draw (C) -- (D);
            \draw (D) -- (E);
            \draw (C) -- (E);

            \draw (F) -- (H);
            \draw (H) -- (J);
            \draw (J) -- (L);
            \draw (F) -- (G);
            \draw (G) -- (H);
            \draw (H) -- (I);
            \draw (I) -- (H);
            \draw (I) -- (J);
            \draw (J) -- (K);
            \draw (K) -- (L);
        \end{tikzpicture}
        \caption{{independent chains}}
        \label{fig:multi-chain-independent}
    \end{subfigure}
    \begin{subfigure}{0.15\textwidth}
        \centering
        \begin{tikzpicture}[scale=0.7]
            \encodingNode{A}{-5.0 + 0.05,0.3 - 0.05}
            \encodingNode{B}{-4.9,1.0}
            \auxNode{C}{-4.3 + 0.05,1.1 - 0.05}
            \encodingNode{D}{-4.2,1.8}
            \auxNode{E}{-3.6 + 0.05,1.9 - 0.05}
            \encodingNode{F}{-3.5,2.6}
            \auxNode{G}{-2.9 + 0.05,2.7 - 0.05}
            \encodingNode{H}{-3.5,1.2}
            \auxNode{I}{-3.1,0.6}
            \encodingNode{J}{-2.3,0.7}
            \auxNode{K}{-1.9,0.1}

            \draw (A) -- (B);
            \draw (A) -- (C);
            \draw (B) -- (C);
            \draw (C) -- (D);
            \draw (D) -- (E);
            \draw (C) -- (E);
            \draw (E) -- (F);
            \draw (F) -- (G);
            \draw (E) -- (G);
            \draw (H) -- (I);
            \draw (H) -- (C);
            \draw (I) -- (C);
            \draw (I) -- (J);
            \draw (I) -- (K);
            \draw (J) -- (K);
        \end{tikzpicture}
        \caption{{chain bifurcations}}
        \label{fig:multi-chain-bifurcation}
    \end{subfigure}
    \begin{subfigure}{0.15\textwidth}
        \centering
        \begin{tikzpicture}[scale=0.7]
            \encodingNode{A}{-5.0 + 0.05,0.3 - 0.05}
            \encodingNode{B}{-4.9,1.0}
            \auxNode{C}{-4.3 + 0.05,1.1 - 0.05}
            \encodingNode{D}{-4.2,1.8}
            \auxNode{E}{-3.6 + 0.05,1.9 - 0.05}
            \encodingNode{F}{-3.5,2.6}
            \auxNode{G}{-2.9 + 0.05,2.7 - 0.05}
            \auxNode{H}{-3.1,1.2}
            \encodingNode{I}{-3.6,0.7}

            \draw (A) -- (B);
            \draw (A) -- (C);
            \draw (B) -- (C);
            \draw (C) -- (D);
            \draw (D) -- (E);
            \draw (C) -- (E);
            \draw (E) -- (F);
            \draw (F) -- (G);
            \draw (E) -- (G);
            \draw (H) -- (E);
            \draw (H) -- (C);
            \draw (H) -- (G);
            \draw (H) -- (I);
            \draw (C) -- (I);
            
        \end{tikzpicture}
        \caption{overlapping chains}
        \label{fig:multi-chain-overlap}
    \end{subfigure}
    \vspace{-0.5em}
    \caption{Potential types of anomalies in the interaction graph if the variable selection algorithm is executed multiple times.}
    \label{fig:multi-chain}
    \vspace{-1.2em}
\end{figure}
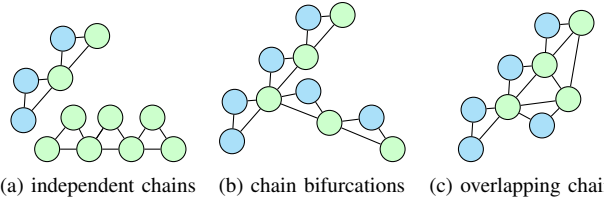

Furthermore, if at some point during the execution, no new~$\bs_{i+1}$ can be found that substitutes a product with the factor $\bs_i$, the algorithm stops early.
In this case, the resulting cost function still contains higher order terms.
To resolve this, the algorithm can be executed again, starting with the cost function $Q(\bx')$, where $\bx' = \bx \cup \bs$.
This repeated execution leads to \emph{additional chains} in the interaction graph which may have different structures, depending on what variables are selected in the subsequent execution of the algorithm.
Examples for each of these altered kinds of structures are illustrated in \autoref{fig:multi-chain}.

\begin{itemize}
    \item \emph{Independent Chains} (\autoref{fig:multi-chain-independent}): If no variable selected for substitution by the second execution was an auxiliary variable or a problem variable selected during the first execution of the variable selection algorithm, the second execution results in two individual chains of triangles in the final interaction graph.
    \item \emph{Chain Bifurcations} (\autoref{fig:multi-chain-bifurcation}): If a variable selected for substitution by the second execution was an auxiliary variable introduced during the first execution, the resulting interaction graph will split into two chains at the corresponding vertex.
    \item \emph{Overlapping Chains} (\autoref{fig:multi-chain-overlap}): If several variables selected for substitution by the second execution were also selected as problem variables or introduced as auxiliary variables in the first execution, the resulting chains of triangles overlap at the corresponding vertices, resulting in a less structured interaction graph.
\end{itemize}

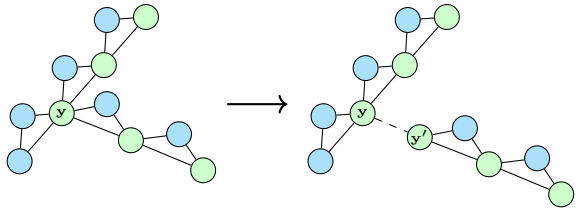
\begin{figure}
\centering
\begin{tikzpicture}[scale=0.8]
    \def\textsize{\tiny}
    \encodingNode{A}{-5.0 + 0.05,0.3 - 0.05}
    \encodingNode{B}{-4.9,1.0}
    \auxNode[$\bs$]{C}{-4.3 + 0.05,1.1 - 0.05}
    \encodingNode{D}{-4.2,1.8}
    \auxNode{E}{-3.6 + 0.05,1.9 - 0.05}
    \encodingNode{F}{-3.5,2.6}
    \auxNode{G}{-2.9 + 0.05,2.7 - 0.05}
    \encodingNode{H}{-3.5,1.2}
    \auxNode{I}{-3.1,0.6}
    \encodingNode{J}{-2.3,0.7}
    \auxNode{K}{-1.9,0.1}

    \draw (A) -- (B);
    \draw (A) -- (C);
    \draw (B) -- (C);
    \draw (C) -- (D);
    \draw (D) -- (E);
    \draw (C) -- (E);
    \draw (E) -- (F);
    \draw (F) -- (G);
    \draw (E) -- (G);
    \draw (H) -- (I);
    \draw (H) -- (C);
    \draw (I) -- (C);
    \draw (I) -- (J);
    \draw (I) -- (K);
    \draw (J) -- (K);

    \def\dx{0.95}
    \def\dy{-0.4}
    \def\gap{5}

    \encodingNode{A'}{\gap + -5.0 + 0.05,0.3 - 0.05}
    \encodingNode{B'}{\gap + -4.9,1.0}
    \auxNode[\textsize $\bs$]{C'}{\gap + -4.3 + 0.05,1.1 - 0.05}
    \auxNode[$\bs'$]{C''}{\gap + -4.3 + 0.05 + \dx,1.1 - 0.05 + \dy}
    \encodingNode{D'}{\gap + -4.2,1.8}
    \auxNode{E'}{\gap + -3.6 + 0.05,1.9 - 0.05}
    \encodingNode{F'}{\gap + -3.5,2.6}
    \auxNode{G'}{\gap + -2.9 + 0.05,2.7 - 0.05}
    \encodingNode{H'}{\gap + -3.5 + \dx,1.2 + \dy}
    \auxNode{I'}{\gap + -3.1 + \dx,0.6 + \dy}
    \encodingNode{J'}{\gap + -2.3 + \dx,0.7 + \dy}
    \auxNode{K'}{\gap + -1.9 + \dx,0.1 + \dy}

    \draw (A') -- (B');
    \draw (A') -- (C');
    \draw (B') -- (C');
    \draw (C') -- (D');
    \draw (D') -- (E');
    \draw (C') -- (E');
    \draw (E') -- (F');
    \draw (F') -- (G');
    \draw (E') -- (G');
    \draw (H') -- (I');
    \draw (H') -- (C'');
    \draw (I') -- (C'');
    \draw (I') -- (J');
    \draw (I') -- (K');
    \draw (J') -- (K');
    \draw[dashed] (C') -- (C'');

    \draw[->, thick] (-1.5, 1.2) -- (-1.5 + 1, 1.2);
\end{tikzpicture}
\vspace{-0.5em}
\caption{The process of splitting a bifurcation into two individual chains by introducing a new variable $\bs'$.}
\label{fig:split-chains}
\vspace{-2.0em}
\end{figure}

Depending on the type of additional chain, the mapping process must be adapted.
In the case of \emph{independent chains}, each chain can be mapped to the target architecture individually.
As no qubits are reused among both chains, all corresponding gates between the two chains can be applied in parallel, leading to no increase in circuit depth.
For \emph{chain bifurcations} or \emph{overlapping chains}, the initial structure must first be modified to transform them into \emph{independent chains}.
This can be achieved by noting that in a binary cost function, the equality $x = x'$ can be enforced by the penalty term~$(x - x')^2$.
Therefore, for each variable $x_i$ shared between two chains, we introduce a new variable $x_i'$. 
We then assign $x_i$ to the first chain and~$x_i'$ to the second chain, adding a single new edge between the two variables due to the penalty term as an \emph{extraneous interaction}.
This process is further illustrated by \autoref{fig:split-chains}.
This once again allows the chains to be handled in parallel in the resulting QAOA circuit.
The total depth is only increased by the cost of inserting \texttt{SWAP} gates for the single newly introduced interaction, while the circuit width is increased depending on the number of shared variables.

\vspace{-0.5em}

\section{Evaluation and Discussion}
\label{sec:evaluation}

To evaluate the proposed methodology, we implemented the core methods with all code publicly available at \redacted{https://github.com/munich-quantum-toolkit/problemsolver} as part of the Munich Quantum Toolkit (MQT,~\cite{mqt}).
We then conducted evaluations, targeting IBM's \texttt{ibm\_torino} superconducting device with 133 qubits for all procedures.

This evaluation focuses on randomly generated cost functions of the form $\sum_i^N c_i \prod \overline{\bx}_i$, where each $\overline{\bx}_i$ is the product of a randomly chosen subset of all $N$ encoding variables $\bx$ and each $c_i$ is a random coefficient with $-10 \leq c_i \leq 10$.
This results in an adequately random selection of cost functions that are not biased towards the proposed solution.
100 inputs were generated for each input size $8 \leq N \leq 16$.

Circuit depth gives valuable insight into the performance of a circuit on NISQ devices. It provides an estimate on both the expected accuracy of the results as well as the expected runtime of the circuit.

\begin{figure}
\centering
\begin{subfigure}{0.24\textwidth}
    \centering
    \includegraphics[width=1\textwidth]{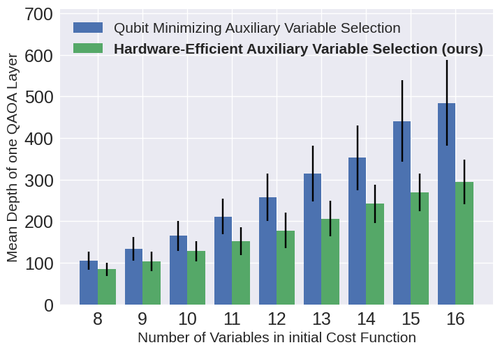}
    \caption{depth}
    \label{fig:eval-size}
\end{subfigure}
\begin{subfigure}{0.24\textwidth}
    \centering
    \includegraphics[width=1\textwidth]{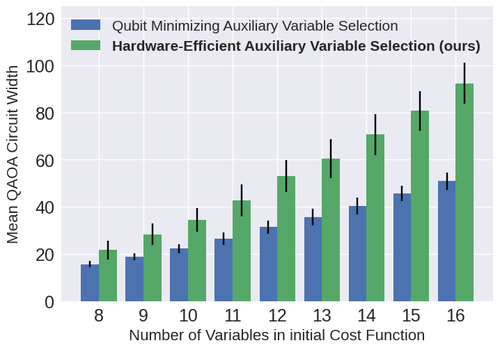}
    \caption{width}
    \label{fig:eval-width}
\end{subfigure}
\vspace{-1.7em}
\caption{The average depth and width of one QAOA repetition layer on \texttt{ibm\_torino}, averaged over 100 samples per size, using the approach proposed in this work compared to a hardware-agnostic strategy that aims to minimize the total number of qubits.}
\vspace{-1.5em}
\end{figure}

Each generated cost function was then compiled to the target architecture using the hardware-efficient auxiliary variable selection proposed in this work.
Simultaneously, using a conventional heuristic to select auxiliary variables in order to minimize the total number of variables, an alternative QUBO formulation was constructed for each input and its corresponding QAOA circuit was mapped to the target device using \emph{Qiskit}'s~\cite{qiskit} \texttt{transpile} method with optimization level 3.
\autoref{fig:eval-size} shows a comprehensive comparison of the resulting circuit depths for both approaches, where a higher value corresponds to a comparatively worse circuit.
It indicates a clear reduction in circuit depth, especially as the size of inputs rises, resulting in an average improvement of 39.2\% at the maximal sizes possible for the 133 qubit device.
\autoref{fig:eval-width} further compares the circuit widths of the approaches.
Here, higher values indicate that the circuit requires a larger number of qubit and is, therefore, more difficult to execute.
As expected, the proposed method for auxiliary variable selection sacrifices width for depth.
For the largest investigated inputs, the conventional auxiliary variable selection method requires an average of 45.0\% fewer qubits than the proposed approach.

\begin{figure}
    \centering
    \includegraphics[width=0.3\textwidth]{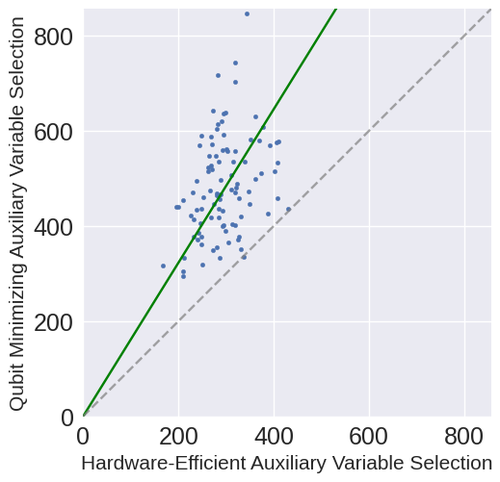}
    \vspace{-0.5em}
    \caption{The depth of a single QAOA repetition layer on \texttt{ibm\_torino} using the approach proposed in this work compared to a hardware-agnostic strategy that aims to minimize the number of qubits for individual test inputs.}
    \label{fig:depth-scatter}
    \vspace{-1.5em}
\end{figure}

Furthermore, \autoref{fig:depth-scatter} shows a detailed comparison between both approaches for individual input functions.
The proposed method consistently performs better in almost all instances and the green trend line through the origin clearly suggests a general improvement as the circuit size increases.

\vspace{-0.5em}
\section{Conclusion}
\label{sec:conclusion}

In this work, we have proposed a novel approach for the selection of auxiliary variables to construct hardware-efficient QUBO formulations.
A greedy algorithm for the selection of substitutions leads to a regular interaction graph structure even as the total number of variables grows.
A secondary algorithm was proposed to map this interaction graph to a QAOA circuit that achieves constant depth in the ideal case.
In cases where the ideal outcome cannot be obtained, we further provided strategies to minimize the resulting limitations.
Evaluations of the proposed method have shown an increase in the advantage gained by this method in terms of circuit depth as the number of variables grows at the expense of circuit width.
Notably, the proposed auxiliary variable selection method is complementary with other compilation techniques.
Even without using the proposed method for QAOA circuit generation, improvements can already be achieved by using conventional compilers with the generated QUBO cost function.
Future work may investigate combining the proposed approach with existing QAOA-optimized compilers to achieve even greater depth reductions.
Furthermore, new strategies, such as qubit reuse, may also be leveraged to reduce the number of required qubits, mitigating the circuit width cost of the proposed methodology.
All proposed methods are made available as part of an open-source implementation at \redacted{https://github.com/munich-quantum-toolkit/problemsolver}.

\redacted[\nothing]{
	\vfill
\section*{Acknowledgments}
\footnotesize This work received funding from the European Research Council (ERC) under the
European Union's Horizon 2020 research and innovation program (grant agreement 
No. 101001318), was part of the Munich Quantum Valley, which is supported by 
the Bavarian state government with funds from the Hightech Agenda Bayern Plus, 
and has been supported by the BMK and BMDW.
We acknowledge the use of IBM Quantum services for this work. 
The views expressed are those of the authors, and do not reflect the official policy or position of IBM or the IBM Quantum team.
}

\clearpage

\printbibliography

\end{document}